\documentclass[11pt,a4paper]{article}
\usepackage{a4wide}
\usepackage{amsmath,amssymb}
\usepackage[mathscr]{eucal}
\newtheorem{theorem}{Theorem}
\newtheorem{definition}{Definition}
\newtheorem{lemma}{Lemma}

\begin{document}
\title{On the existence of Lyapounov variables for Schr\"odinger evolution}
\author{Y. Strauss\thanks{E-mail: ystrauss@math.bgu.ac.il, yossef.strauss@gmail.com}\\
            Department of Mathematics\\
            Ben Gurion University of the Negev\\
            Be'er Sheva 84105, Israel}
\maketitle
\begin{abstract}
The theory of (classical and) quantum mechanical microscopic irreversibility developed by B. Misra, I. Prigogine and M. Courbage
(MPC) and various other contributors is based on the central notion of a Lyapounov variable - i.e., a dynamical variable whose
value varies monotonically as time increases. Incompatibility between certain assumed properties of a Lyapounov variable and
semiboundedness of the spectrum of the Hamiltonian generating the quantum dynamics led MPC to formulate their theory in 
Liouville space. In the present paper it is proved, in a constructive way, that a Lyapounov variable can be found within the standard
Hilbert space formulation of quantum mechanics and, hence, the MPC assumptions are more restrictive than necessary for the
construction of such a quantity. Moreover, as in the MPC theory, the existence of a Lyapounov variable implies the existence of 
a transformation (the so called $\Lambda$-transformation) mapping the original quantum mechanical problem to an equivalent 
irreversible representation. In addition, it is proved that in the irreversible representation there exists a natural time observable
splitting the Hilbert space at each $t>0$ into past and future subspaces.
\end{abstract}
\section{introduction}
\label{introduction}
\par During the late 1970's and in the following decades a comprehensive theory of classical and quantum microscopic
irreversibility has been developed by B. Misra, I. Prigogine and M. Courbage and various other contributors (see for exmple 
\cite{M,MPC1,MPC2,Pr,C,GGM,AM} and references therein). A central notion in this theory of 
irreversibility is that of a non-equilibrium entropy associated with the existence of Lyapounov variables for the dynamical system
under consideration. In the case of a classical dynamical system one works with Koopman's formulation of classical mechanics in
Hilbert space \cite{Kp}. Associated with the dynamical system there exists a measure space $(\Omega,\mathcal F,\mu)$ such 
that $\Omega$ consists of all points belonging to a constant energy surface in phase space, $\mathcal F$ is a $\sigma$-algebra 
of measurable sets with respect to the measure $\mu$ which is taken to be the Liouville measure invariant under the Hamiltonian
evolution. The Hamiltonian dynamics is given in terms of a one parameter dynamical group $T_t$ mapping $\Omega$ onto itself
with the condition that, for all $t$, $T_t$ is measure preserving and injective.
The Koopman Hilbert space is then the space $\mathcal H=L^2(\Omega,d\mu)$ of functions on $\Omega$ square 
integrable with respect to $\mu$. The dynamics of the system is represented in $\mathcal H$ by a one-parameter unitary group 
$\{U_t\}_{t\in\mathbb R}$ induced by the group $T_t$ via 
\begin{equation*}
 (U_t\psi)(\omega)=\psi(T_{-t}\,\omega),\qquad \omega\in\Omega,\ \psi\in\mathcal H\,.
\end{equation*}
The generator of $\{U_t\}_{t\in\mathbb R}$ is the Liouvillian $L$ 
\begin{equation*}
 U_t=e^{-iLt},\qquad t\in\mathbb R\,.
\end{equation*}
The generator $L$ is, in general, an unbounded self-adjoint operator. For a Hamiltonian system it is given by
\begin{equation}
\label{liouvillian_hamilton_sys}
 L\psi=i[H,\psi]_{pb}\,,
\end{equation}
where the subscript $pb$ stands for Poisson brackets and where Eq. (\ref{liouvillian_hamilton_sys}) holds for all $\psi$ for which its
right hand side is well defined in $\mathcal H$.
\par A bounded, non-negative, self-adjoint operator $M$ in $L^2(\Omega,d\mu)$ is called a \emph{Lyapounov variable} essentially if
it satisfies the condition that, for every $\psi\in\mathcal  H$ the quantity
\begin{equation}
\label{lyapounov}
 (\psi_t,M\psi_t)=(U_t\psi,MU_t\psi)
\end{equation}
is a monotonically decreasing function of $t$. This monotonicity property of $M$ allows, within the framework of the theory of
irreversibity mentioned above, to define the notion of non-equilibrium entropy and the second law of thermodynamics as a 
fundamental dynamical principle. Of course, the monotonicity condition is not sufficient for the definition of non-equilibirium entropy 
and further conditions are introduced. These conditions will be discussed below.
\par In reference \cite{M} a Lyapounov variable for a classical system is defined as a bounded operator $M$ in 
$L^2(\Omega,d\mu)$ satisfying the conditions
\begin{enumerate}
\item $M$ is a non-negative operator.
\item $\mathcal D(L)$, the domain of  $L$, is stable under the action of $M$, i.e., $M\mathcal D(L)\subseteq\mathcal D(L)$.
\item $i[L,M]\subseteq -D$ where $D$ is a non-negative, self-adjoint operator in $\mathcal H$.
\item $(\psi, D\psi)=0$ iff $\psi(\omega)=const.$, $\omega\in\Omega$, a.e on $\Omega$.
\end{enumerate}
It is remarked in reference \cite{M} that if, for a bounded operator $M$ on $L^2(\Omega,d\mu)$, the quantity $(\psi_t,M\psi_t)$
has the required monotonicity property then $M$ may be considered as a Lyapounov variable except for the fact that condition 
(2) above on the stability of $\mathcal D(L)$ may not be satisfied. In this respect an important observation for our purposes is
that instability of $\mathcal D(L)$ under the action of $M$ has direct consequences on the domain of definition of the commutator
$i[L,M]$ appearing in condition (3).
\par In Koopman's Hilbert space formulation of classical mechanics all physical observables of the classical system have a natural
represention as mutiliplicative operators in $L^2(\Omega,d\mu)$. By a theorem of Poincar\'e \cite{Po} there is no function on
phase space that has a
definite sign and is monotonically increasing under the Hamiltonian evolution. This leads to the conclusion that non-equilibrium
entropy, or a Lyapounov variable, cannot be represented as a mutliplicative operator in the corresponding Hilbert space formulation
of classical mechanics \cite{M}. In fact, a Lyapounov variable $M$ does not commute with at least some of the operators of 
multiplication by phase space functions.
\par Consider now the framework of quantum mechanics where a physical system is described by a Hilbert space $\mathcal H$
and the quantum mechanical evolution is generated by a self-adjoint Hamiltonian operator $H$. Let $\mathcal B(\mathcal H)$ be the
space of bounded operators defined on $\mathcal H$ and let 
$M\in\mathcal B(\mathcal H)$ be an operator in $\mathcal H$ representing a Lyapounov variable (corresponding to non-equilibrium
entropy) and assume that 
\begin{enumerate}
\item[i)] $H$ is bounded from below; with no restriction of generality we may assume that $H\geq 0$,
\item[ii)] $M$ is self-adjoint,
\item[iii)] $\mathcal D(H)$, the domain of $H$, is stable under the action of $M$,
\item[iv)] $i[H,M]\subseteq -D$ where $D$ is self-adjoint on $\mathcal H$ and $D\geq 0$,
\item[v)] $[M,D]=0$
\end{enumerate}
\bigskip
\par{\bf Remarks:} Condition (v) is to be interpreted as meaning that $DM$ extends $MD$, i.e., $MD\subset DM$. In addition 
condition (iii) implies that the commutator $i[H,M]$ in condition (iv) is defined on $\mathcal D(H)$. Condition (iv) then
implies that $\mathcal D(D)\supseteq\mathcal D(H)$.
\bigskip
\par\noindent Under a set of assumptions equivalent to conditions (i)-(v) above it has been proven by Misra, Prigogine and 
Courbage \cite{MPC1} that ptions (i)-(v) that $D\equiv 0$ and hence $M$ cannot be a Lyapounov variable. The crucial element of
the proof is the fact that $H$ is bounded from below. The solution found by the authors of reference \cite{MPC1} is to work with
the Liouvillian formulation of quantum mechanics where the quantum evolution is acting on the space of density operators and the
generator of evolution is the Liouvillian $L$ defined by
\begin{equation*}
 L\rho=[H,\rho]
\end{equation*}
with $\rho$ a density operator. For example, if the Hamiltonian $H$ satisfies the condition that 
$\sigma(H)=\sigma_{ac}(H)=\mathbb R^+$ then the Liouvillian $L$ has an absolutely continuous spectrum of
uniform (infinite) multiplicity consisting of all of $\mathbb R$. It is then possible to avoid the conclusions of the 
Poincare'-Misra no-go theorem and define $M$ as a superoperator on the space of density operators satisfying in this space the
conditions
\begin{equation*}
 i[L,M]\subseteq -D\leq 0,\qquad [M,D]=0\,.
\end{equation*}
As mentioned above the monotnicity condition in the form of the existence of a Lyapounov variable $M$ is not enough to 
identify $M$ as an operator (in the classical case) or a superoperator (in the quantum case) representing non-equilibrium entropy.
If $M$ is a Lyapounov variable corresponding to non-equilibrium entropy one would also like to be able to use it in order to describe
the process of decay of deviations from equilibrium for the physical system under consideration and recover the unidirectional nature
of the evolution of such a system. A theory of transformation, via non-unitary mappings, between conservative dynamics 
represented by unitary evolution $\{U_t\}_{t\in\mathbb R}$ and dissipative dynamics represented by semigroup evolution
$\{W_t\}_{t\in\mathbb R^+}$ has been developed by Misra, Prigogine and Courbage \cite{MPC2,CM,GMC,MP,Pr,GGM} for
systems with internal time operator $T$ satisfying
\begin{equation}
\label{internal_t_eqn}
 U_{-t}TU_t=T+tI\,.
\end{equation}
In this formalism the time operator $T$ is canonically conjugate to the generator $L$ of the unitary evolution group of the 
conservative dynamics, i.e. $U_t=\exp(-iLt)$ and 
\begin{equation}
\label{L_T_canon_conj_eqn}
 [L,T]=i\,.
\end{equation}
For a system possessing internal time operator $T$ it is possible to construct a Lyapounov variable as a positive monotonically 
decreasing operator function $M=M(T)$. One is then able to define a non-unitary transformation 
\begin{equation}
\label{lambda_transform_eqn}
 \Lambda=\Lambda(T)=(M(T))^{1/2}
\end{equation}
such that 
\begin{equation}
\label{unitary_semigroup_intertwine_eqn}
 \Lambda U_t=W_t\Lambda,\qquad t\geq 0
\end{equation}
where, as in the discussion above, $W_t$ is a dissipative semigroup with  
\begin{equation*}
 \Vert W_{t_2}\psi\Vert\leq\Vert W_{t_1}\psi\Vert,\qquad t_2\geq t_1,\ \ \psi\in\mathcal H
\end{equation*}
and 
\begin{equation*}
 s-\lim_{t\to\infty} W_t=0\,.
\end{equation*}
Note that Eq. (\ref{L_T_canon_conj_eqn}) implies that $\sigma(L)=\mathbb R$ and $\sigma(T)=\mathbb R$ so that, in 
order for the Misra, Prigogine and Courbage formalism to work, it is required that the generator of evolution is unbounded from
below. 
\par The goal of the present paper is to show that it is possible to define the main objects and obtain many of the results of the
Misra, Prigogine and Courbage theory within the standard formulation of quantum theory without ever invoking the need to work in a
more generalized space such as Liouville space. It will be shown in Theorem \ref{exist_lyapounov_var_thm} below
that, under the same assumptions on the spectrum of the Hamiltonian as in the Misra, Prigogine
and Courbage theory, the semiboundedness of the Hamiltonian does not hinder the possiblity of defining a Lyapounov variable for
the Schr\"odinger evolution. Of course, in light of the no-go theorem discussed above, at least one of the conditions (i)-(v) above
is not satisfied in the construction of this Lyapounov variable and, in fact, conditions (iii),(iv) and (v) do not hold in this 
construction. Detailed discussion of this point will be given elsewhere. 
\par Theorem \ref{exist_lyapounov_var_thm} concerning the existence of a Lyapounov variable $M_F$ for forward (positive times)
Schr\"odinger evolution is stated at the beginning of Section \ref{main_results_sec} and proved in Section \ref{proofs}. 
\par Once the existence of the Lyapounov variable $M_F$ is established one can proceed as in the Misra, Prigogine and Courbage 
theory and define a non-unitary $\Lambda$-transformation $\Lambda_F=M_F^{1/2}$ as in Eq. (\ref{lambda_transform_eqn}). It is
to be emphasized, however, that the existence of a time operator satisfying Eq. (\ref{L_T_canon_conj_eqn}) is not required in the
construction of $M_F$ and $\Lambda_F$ and, in fact, since the spectrum of $H$ is bounded from below, such a time operator
does not exist.
\par It is shown in Section \ref{main_results_sec} that via the $\Lambda$-transformation $\Lambda_F$ it is possible to establish
a relation of the form given in Eq. (\ref{unitary_semigroup_intertwine_eqn}), i.e., there exists a dissipative one-parameter
continuous semigroup $\{ Z(t)\}_{t\in\mathbb R^+}$ such that for $t\geq 0$ we have
\begin{equation*}
 \Lambda_F U(t)=Z(t)\Lambda_F,\quad t\geq 0\,,
\end{equation*}
one is then able to obtain the \emph{irreversible representation} of the dynamics. This is done in Section \ref{main_results_sec}.
In the irreversible representation the dynamics of the system is unidirectional in time and is given in terms of the semigroup 
$\{Z(t)\}_{t\in\mathbb R^+}$.
\par It is an interesting fact that in the irreversible representation of the dynamics it is possible to find a positive 
semibounded operator $T$ in $\mathcal H$ that can be intepreted as a natural time observable for the evolution of the system.
The exact nature of this time observable and the main theorem concerned with its existence is discussed in 
Section \ref{main_results_sec}. Of course, this operator is not a time operator in the sense of Eq. (\ref{internal_t_eqn}).
\par Following a statement in Section \ref{main_results_sec} of the main theorems proved in this paper and the discussion of
their content, the proofs of these theorems are provided in Section \ref{proofs}. A short summary is provided in 
Section \ref{summary}.
\section{Main theorems and results}
\label{main_results_sec}
\par The three main theorems in this section, and the discussion accompanying them, provide the main results of the present
paper. We start with the existence of Lyapounov variables for Schr\"odinger evolution.
\begin{theorem}
\label{exist_lyapounov_var_thm}
Assume that:
\begin{description}
\item[a)] $\mathcal H$ is a separable Hilbert space and $\{U(t)\}_{t\in\mathbb R}$ is a unitary evolution group defined on
$\mathcal H$,
\item[b)] the generator $H$ of $\{U(t)\}_{t\in\mathbb R}$ is self-adjoint on a dense domain 
$\mathscr D(H)\subset\mathcal H$ and $\sigma(H)=\sigma_{ac}(H)=\mathbb R^+$,
\item[c)] The spectrum $\sigma(H)$ is of multiplicity one (see remark below).
\end{description}
Let $\{\phi_E\}_{E\in\mathbb R^+}$ be a complete set of improper eigenvectors of $H$ corrsponding to the spectrum of 
$H$. We shall use the Dirac notation and denote $\{\vert E\rangle\}_{E\in\mathbb R^+}\equiv\{\phi_E\}_{E\in\mathbb R^+}$.
Then, under the assumptions (a)-(c) above there exists a self-adjoint, contractive, injective, non-negative operator 
$M_F:\mathcal H\mapsto\mathcal H$ 
\begin{equation}
\label{M_F_expression_eqn}
 M_F=M_F(H)=\frac{-1}{2\pi i}\int_0^\infty dE'\int_0^\infty dE\, \vert E'\rangle\,\frac{1}{E'-E+i0^+}\,\langle E\vert
\end{equation}
such that $Ran\,M_F\subset\mathcal H$ is dense in $\mathcal H$ and 
$M_F$ is a Lyapounov variable for the Schr\"odinger evolution in the forward direction, i.e., for every $\psi\in\mathcal H$ we have
\begin{equation}
\label{M_F_lyapounov_var_eqn}
 (\psi_{t_2},M_F\,\psi_{t_2})\leq (\psi_{t_1},M_F\,\psi_{t_1}),\qquad t_2>t_1\geq 0,\ \ \psi_t=U(t)\psi=e^{-iH t}\psi,\ t\geq 0
\end{equation}
and
\begin{equation}
\label{M_F_to_zero_eqn}
 \lim_{t\to\infty} (\psi_t,M_F\,\psi_t)=0\,.
\end{equation}
\hfill$\square$
\end{theorem}
{\bf Remark:} Assumption (c) above is made for simplicity of proof and exposition. The result has immediate generalization to a
spectrum of any finite multiplicity. The case of infinite multiplicity will be cosidered separately elsewhere.
\bigskip
\par Following the proof of the existence of the Lyapounov variable $M_F$ we can proceed as in the Misra, Prigogine and Courbage
theory and obtain a non-unitary $\Lambda$ transformation via the definition $\Lambda_F:=M_F^{1/2}$. We then have the following
theorem
\begin{theorem}
\label{lambda_transform_thm}
Let $\Lambda_F:=M_F^{1/2}$. Then $\Lambda_F:\mathcal H\mapsto\mathcal H$ is positive, contractive and quasi-affine map, i.e., 
it is a positive, contractive, injective operator such that $Ran\,\Lambda_F$ is dense in $\mathcal H$. Furthermore, there exists a
continuous, strongly contractive, one parameter semigroup $\{Z(t)\}_{t\in\mathbb R^+}$ such that
\begin{equation}
\label{Z_semigroup_prop_eqn}
 \Vert Z(t_2)\psi\Vert\leq \Vert Z(t_1)\psi\Vert,\ \ t_2\geq t_1\geq 0,\qquad 
 s-\lim_{t\to\infty} Z(t)=0\,.
\end{equation}
and the following intertwining relation holds
\begin{equation}
\label{Z_intertwining_eqn}
 \Lambda_F U(t)=Z(t)\Lambda_F,\qquad U(t)=e^{-i Ht},\ \ t\geq 0\,.
\end{equation}
\hfill$\square$
\end{theorem}
Taking the adjoint of Eq. (\ref{Z_intertwining_eqn}) we obtain the intertwining relation
\begin{equation}
\label{Z_intertwining_adj_eqn}
 U(-t)\Lambda_F=\Lambda_F Z^*(t),\quad t\geq 0\,.
\end{equation}
Let $\mathscr L(\mathcal H)$ be the set of linear operators in $\mathcal H$. For $X\in\mathscr L(\mathcal H)$ denote 
$X_{\Lambda_F}:=\Lambda_F X\Lambda_F$ and consider the set of all self-adjoint operators $X\in\mathscr L(\mathcal H)$ such
that $\mathscr D(X_{\Lambda_F})$ is dense in $\mathcal H$ i.e., such that $\Lambda_F^{-1}\mathscr D(X)$ is dense in 
$\mathcal H$. Using Eqns. (\ref{Z_intertwining_eqn}) and (\ref{Z_intertwining_adj_eqn}) we obtain
\begin{equation}
\label{heisenberg_to_semigroup_eqn}
 U(-t)X_{\Lambda_F} U(t)=
 U(-t)\Lambda_F X\Lambda_F U(t)=\Lambda_F Z^*(t) X Z(t)\Lambda_F,\quad t\geq 0\,.
\end{equation}
For $\varphi,\psi\in\mathcal H$ denote $\varphi_{\Lambda_F}=\Lambda_F\varphi$ and $\psi_{\Lambda_F}=\Lambda_F\psi$.
Then using Eq. (\ref{heisenberg_to_semigroup_eqn}) implies that
\begin{equation}
\label{QM_semigroup_rep_eqn}
 (\varphi,U(-t)X_{\Lambda_F}U(t)\psi)=(\varphi,\Lambda_FZ^*(t)XZ(t)\Lambda_F\psi)
 =(\varphi_{\Lambda_F},Z^*(t)XZ(t)\psi_{\Lambda_F}),\quad t\geq 0\,.
\end{equation}
We shall assume that \emph{each and every relevant physical observable of the original problem has a representation in the form
$X_{\Lambda_F}=\Lambda_F X\Lambda_F$ for some self-adjoint $X\in\mathscr L(\mathcal H)$}. Then, since the left
hand side of Eq. (\ref{QM_semigroup_rep_eqn}) corresponds to the original quantum mechanical problem, the right hand side of this
equation constitutes a new representation of the original problem in terms of the correspondence
\begin{eqnarray*}
 \psi &\longrightarrow& \psi_{\Lambda_F}=\Lambda_F\psi\\
 U(t) &\longrightarrow& Z(t)=\Lambda_F\,U(t)\Lambda_F^{-1},\ \ t\geq 0\\
 X_{\Lambda_F} &\longrightarrow&\ \ \ X=\Lambda_F^{-1}X_{\Lambda_F}\Lambda_F^{-1}\,.
\end{eqnarray*}
Considering the fact that on the right hand side of Eq. (\ref{QM_semigroup_rep_eqn}) the dynamics is given in terms of the 
semigroup $\{Z(t)\}_{t\in\mathbb R^+}$  we may call the right hand side of 
Eq. (\ref{QM_semigroup_rep_eqn}) the \emph{irreversible representation} of the problem. The left hand side of that equation is
then the \emph{reversible representation} (or the standard representation). 
\par It is an interesting fact that in the irreversible representation of a quantum mechanical problem, as in the right hand side
of Eq. (\ref{QM_semigroup_rep_eqn}), one can find a self-adjoint operator $T$ with continuous spectrum $\sigma(T)=([0,\infty))$ 
such that for every $t\geq 0$ the spectral projections on the spectrum of $T$ naturally divide the Hilbert space $\mathcal H$ into a
direct sum of a \emph{past subspace at time $t$} and a \emph{future subspace at time $t$}. Specifically, we have the following 
theorem:
\begin{theorem}
\label{T_observable_thm}
Let $\mathscr B(\mathbb R^+)$ be the Borel $\sigma$-algebra generated by open subsets of $\mathbb R^+$ and 
$\mathcal P(\mathcal H)$ be the set of orthogonal projections in $\mathcal H$. There exists a semi-bounded, 
self-adjoint operator $T:\mathscr D(T)\mapsto\mathcal H$ defined on a dense domain 
$\mathscr D(T)\subset\mathcal H$ with continuous spectrum $\sigma(T)=[0,\infty)$ and corresponding spectral measure 
$\mu_T :\mathscr B(\mathbb R^+)\mapsto \mathcal P(\mathcal H)$ such that for each $t\geq 0$ 
\begin{equation*}
 \mu_T([0,t])\mathcal H=[Z(t),Z^*(t)]\mathcal H=Ker\, Z(t),\qquad t\geq 0
\end{equation*}
and
\begin{equation*}
 \mu_T([t,\infty))\mathcal H=Z^*(t)Z(t)\mathcal H=(Ker\,Z(t))^\perp,\qquad t\geq 0\,.
\end{equation*}
In particular, for $0<t_1< t_2$ we have $Ker\,Z(t_1)\subset Ker\,Z(t_2)$. For $t=0$ we have $Ker\,Z(0)=\{0\}$ and finally
$\lim_{t\to\infty} Ker\,Z(t)=\mathcal H$. \hfill$\square$
\end{theorem}
Denote the orthogonal projection on $Ker\,Z(t)$ by $P_{t]}$ and the orthogonal projection on $(Ker\,Z(t))^\perp$ by $P_{[t}$\ . From
Theorem \ref{T_observable_thm} we have for $t\geq 0$
\begin{equation*}
 P_{t]}=[Z(t),Z^*(t)],\qquad P_{[t}=Z^*(t)Z(t),\ \ t\geq 0\,.
\end{equation*}
The projection $P_{t]}$ will be called below the \emph{projection on the past subspace at time $t$}. The projection $P_{[t}$ will be
called the \emph{projection on the future subspace at time $t$}. In accordance we will call $\mathcal H_{t]}:=Ran\,P_{t]}=Ker\,Z(t)$
the \emph{past subspace at time $t$} and $\mathcal H_{[t}:=Ran\,P_{[t}=(Ker\,Z(t))^\perp$ the 
\emph{future subspace at time $t$}. The origin of
the terminology used here can be found in Eq. (\ref{QM_semigroup_rep_eqn}). Using the notation for the projection on $Ker\,Z(t)$
we observe that this equation may be written in the form
\begin{equation*}
 (\varphi,U(-t)X_{\Lambda_F}U(t)\psi)=(P_{[t}\,\varphi_{\Lambda_F},Z^*(t)XZ(t)P_{[t}\,\psi_{\Lambda_F}),\quad t\geq 0\,,
\end{equation*}
and denoting $\varphi_{\Lambda_F}^+(t):= P_{[t}\,\varphi_{\Lambda_F}=P_{[t}\Lambda_F\varphi$ and
$\psi_{\Lambda_F}^+(t):= P_{[t}\,\psi_{\Lambda_F}=P_{[t}\Lambda_F\psi$ we can write in short
\begin{equation}
\label{QM_semigroup_rep_B_eqn}
 (\varphi,U(-t)X_{\Lambda_F}U(t)\psi)=(\varphi^+_{\Lambda_F}(t),Z^*(t)XZ(t)\psi^+_{\Lambda_F}(t)),\quad t\geq 0\,.
\end{equation}
Note that in the irreversible representation on the right hand side of Eq. (\ref{QM_semigroup_rep_B_eqn}) only the projection
of $\varphi_{\Lambda_F}$ and $\psi_{\Lambda_F}$ on the future subspace $\mathcal H_{[t}$ at time $t$ is relevant for the 
calculation of all matrix elements and expectation values for times $t'\geq t\geq 0$. In other words, at time $t$ the subspace
$\mathcal H_{t]}=P_{t]}\mathcal H$ already belongs to the past and is irrelevant for calculations related to the future evolution of the
system. We see that in the irreversible representation the spectral projections of the operator $T$ provide the time ordering of the
evolution of the system. Following these observations it is natural to call $T$ a \emph{time observable} for the irreversible
representation. Note, in particular, that since $M_F=\Lambda_F^2$ we have $\Lambda_F^{-1}M_F\Lambda_F^{-1}=I$ and if we 
plug this relation in Eq. (\ref{QM_semigroup_rep_eqn}) or Eq. (\ref{QM_semigroup_rep_B_eqn}) and take $\varphi=\psi$ we obtain
\begin{multline*}
 (\psi_t,M_F\,\psi_t)=(\psi,U(-t)M_FU(t)\psi)=(\psi_{\Lambda_F},Z^*(t)Z(t)\psi_{\Lambda_F})=\\
 =(\psi_{\Lambda_F},P_{[t}\,\psi_{\Lambda_F})=(\psi_{\Lambda_F},\mu_T([t,\infty))\psi_{\Lambda_F}),\quad t\geq 0\,,
\end{multline*}
thus we have direct correspondence between the Lyapounov variable $M_F$ in the reversible representation of the problem
and the time observable in the irreversible representation. 
\section{Proofs of Main results}
\label{proofs}
\par The basic mechanism underlying the proofs of Theorem \ref{exist_lyapounov_var_thm} and 
Theorem \ref{lambda_transform_thm} is a fundamental intertwining relation, via a quasi-affine mapping, between the unitary 
Schr\"odinger evolution in physical space $\mathcal H$ and semigroup evolution in Hardy space of the upper half-plane 
$\mathcal H^2(\mathbb C^+)$ or the isomorphic space $\mathcal H^2_+(\mathbb R)$ of boundary values on $\mathbb R$ of 
functions in $\mathcal H^2_+(\mathbb C)$. Hence we begin our proof with a few facts concerning Hardy space functions which are
used below. 
\par Denote by $\mathbb C^+$ the upper half of the complex plane. The Hardy space $\mathcal H^2(\mathbb C^+)$ of the upper 
half-plane consists of functions analytic in $\mathbb C^+$ and satisfying the condition that for any 
$f\in\mathcal H^2(\mathbb C^+)$ there exists a constant $C_f>0$ such that
\begin{equation*}
 \sup_{y>0}\int_{-\infty}^{\infty} dx\,\vert f(x+i y)\vert^2< C_f\,.
\end{equation*}
In a similar manner the Hardy space $\mathcal H^2(\mathbb C^-)$ consists of functions analytic in the lower half-plane 
$\mathbb C^-$ and satisfying the condition that for any $g\in\mathcal H^2(\mathbb C^-)$ there exists a constant $C_g>0$ such that
\begin{equation*}
 \sup_{y>0}\int_{-\infty}^{\infty} dx\,\vert f(x-i y)\vert^2< C_g\,.
\end{equation*}
Hardy space functions have non-tangential boundary values a.e. on $\mathbb R$. In particular, for $f\in\mathcal H^2(\mathbb C^+)$
there exists a function $\tilde f\in L^2(\mathbb R)$ such that a.e on $\mathbb R$ we have
\begin{equation*}
 \lim_{y\to 0^+} f(x+iy)=\tilde f(x),\quad x\in\mathbb R\,.
\end{equation*}
a similar limit from below the real axis holds for functions in $\mathcal H^2(\mathbb C^-)$. In fact $\mathcal H^2(\mathbb C^\pm)$
are Hilbert spaces with scalar product given by
\begin{equation*}
 (f,g)_{\mathcal H^2(\mathbb C^\pm)}=\lim_{y\to 0^+}\int_{-\infty}^{\infty} dx\, \overline{f(x\pm i y)}g(x\pm i y)
 =\int_{-\infty}^{\infty} dx\, \overline{\tilde f(x)}\tilde g(x),\quad f,g\in\mathcal H^2(\mathbb C^\pm)\,,
\end{equation*}
where $\tilde f$, $\tilde g$ are the boundary value functions of $f$ and $g$ respectively. The spaces of boundry
values on $\mathbb R$ of functions in $\mathcal H^2(\mathbb C^\pm)$ are then Hilbert spaces isomorphic to 
$\mathcal H^2(\mathbb C^\pm)$ which we denote by $\mathcal H^2_\pm(\mathbb R)$.
\par A Theorem of Titchmarsh \cite{T} states that Hardy space functions can be reconstructed from their boundary value 
functions. If $\tilde f_\pm\in\mathcal H^2_\pm(\mathbb R)$ is a boundary value function of a function 
$f\in\mathcal H^2(\mathbb C^\pm)$ then one has
\begin{equation}
\label{titchmarsh_thm_eqn}
 f(z)=\mp\frac{1}{2\pi i}\int_{-\infty}^\infty dx\,\frac{\tilde f(x)}{z-x}
\end{equation}
where the minus sign corresponds to functions in $\mathcal H^2(\mathbb C^+)$ and the plus sign corresponds to functions in
$\mathcal H^2(\mathbb C^-)$. In addition we shall make use below of the fact that 
\begin{equation*}
 \mathcal H^2_+(\mathbb R)\oplus\mathcal H^2_-(\mathbb R)=L^2(\mathbb R)\,.
\end{equation*}
It can be shown that for functions in $\mathcal H^2(\mathbb C^\pm)$ have radial limits of order $o(z^{-1/2})$ as $\vert z\vert$ goes 
to infinity in the upper and lower half-plane repectively. As a consequence, if we denote by $P_+$ and $P_-$ the projections of
$L^2(\mathbb R)$ on $\mathcal H^2_+(\mathbb R)$ and $\mathcal H^2_-(\mathbb R)$ respectively, Eq. (\ref{titchmarsh_thm_eqn})
and the existence of boundary value functions in $\mathcal H^2_\pm(\mathbb R)$ provides us with explicit expressions for these
projections in the form
\begin{equation}
\label{hardy_projections_eqn}
 (P_\pm f)(\sigma')=\pm\frac{1}{2\pi i}\int_{-\infty}^\infty d\sigma\,\frac{1}{\sigma'-\sigma+i0^+}f(\sigma),\qquad f\in L^2(\mathbb R),
 \ \ \sigma'\in\mathbb R\,.
\end{equation}
The literature on Hardy spaces is quite rich. Additional important properties of Hardy spaces can be found
in \cite{Ks,D,H}. For the vector valued case see, for example, \cite{RR}.
\par Define a family $\{u(t)\}_{t\in\mathbb R}$ of unitary multiplicative operators $u(t): L^2(\mathbb R)\mapsto L^2(\mathbb R)$
by
\begin{equation*}
 [u(t)f](\sigma)=e^{-i\sigma t}f(\sigma),\qquad f\in L^2(\mathbb R),\ \ \sigma\in\mathbb R\,.
\end{equation*}
The family $\{u(t)\}_{t\in\mathbb R}$ forms a one parameter group of multiplicative operators in $L^2(\mathbb R)$. Let $P_+$ be 
the orthogonal projection of $L^2(\mathbb R)$ on $\mathcal H^2_+(\mathbb R)$. A \emph{Toeplitz operator} with symbol 
$u(t)$ \cite{RR,N1,N2} is an operator $T_u(t):\mathcal H^2_+(\mathbb R)\mapsto\mathcal H^2_+(\mathbb R)$ defined by
\begin{equation*}
 T_u(t)f=P_+u(t)f,\qquad f\in\mathcal H^2_+(\mathbb R)\,.
\end{equation*}
The set $\{T_u(t)\}_{t\in\mathbb R^+}$ forms a strongly continuous, contractive, one parameter semigroup on 
$\mathcal H^2 _+(\mathbb R)$ satisfying
\begin{equation}
\label{T_semigroup_eqn}
 \Vert T_u(t_2)f\Vert \leq \Vert T_u(t_1)f\Vert,\qquad t_2\geq t_1\geq 0,\ \ f\in\mathcal H^2_+(\mathbb R)\,,
\end{equation}
and 
\begin{equation}
\label{T_to_zero_eqn}
 s-\lim_{t\to\infty} T_u(t)=0\,.
\end{equation}
Below we shall make frequent use of quasi-affine mappings. The definition of this class of maps is as follows:
\begin{definition}[quasi-affine map]
A quasi-affine map from a Hilbert space $\mathcal H_1$ into a Hilbert space $\mathcal H_0$ is a linear, injective, continuous 
mapping of $\mathcal H_1$ into a dense linear manifold in $\mathcal H_0$. If $A\in\mathscr B(\mathcal H_1)$ and 
$B\in\mathscr B(\mathcal  H_0)$ then $A$ is a quasi-affine transform of $B$ if there is a quasi-affine map $\theta:\mathcal 
H_1\mapsto \mathcal H_0$ such that $\theta A=B\theta$. \hfill$\square$
\end{definition}
Concerning quasi-affine maps we have the following two important facts (see, for example \cite{NF}):
\begin{description}
\item[I)] If $\theta:\mathcal H_1\mapsto\mathcal H_0$ is a quasi-affine mapping then $\theta^*:\mathcal H_0\mapsto\mathcal H_1$
is also quasi-affine, that is, $\theta^*$ is one to one, continuous and its range is dense in $\mathcal H_1$.
\item[II)] If $\theta_1:\mathcal H_0\mapsto\mathcal H_1$ is quasi-affine and $\theta_2:\mathcal H_1\mapsto\mathcal H_2$ is
quasi-affine then $\theta_2\theta_1:\mathcal H_0\mapsto\mathcal H_2$ is quasi-affine.
\end{description}
We can now turn to the proof of Theorem \ref{exist_lyapounov_var_thm}:
\smallskip
\par{\bf Proof of Theorem \ref{exist_lyapounov_var_thm}:}
\smallskip
\par Assume that (a)-(c) in the statement of Theorem \ref{exist_lyapounov_var_thm} hold. A slight variation of a theorem first
proved in \cite{S1}, and subsequently used in the study of resonances in \cite{S1,S2,SHV} and time observables in quantum 
mechanics in \cite{S3}, states that there exists a mapping $\Omega_f:\mathcal H\mapsto\mathcal H^2_+(\mathbb R)$ such that 
\begin{description}
\item[$\alpha$)] $\Omega_f$ is a contractive quasi-affine mapping of $\mathcal H$ into $\mathcal H^2_+(\mathbb R)$.
\item[$\beta$)]   For $t\geq 0$, the Schr\"odinger evolution $U(t)$ is a quasi-affine transform of the Toeplitz operator 
$T_u(t)$. For every $t\geq 0$ and $g\in\mathcal H$ we have
\begin{equation}
\label{basic_intertwining_rel_eqn}
 \Omega_fU(t)g=T_u(t)\Omega_f g,\qquad t\geq 0,\ \ g\in\mathcal H\,.
\end{equation}
\end{description}
(here the subscript $f$ in $\Omega_f$ designates forward time evolution). By (I) above the adjoint 
$\Omega_f^*:\mathcal H^2_+(\mathbb R)\mapsto\mathcal H$ is a quasi-affine map. Hence,
$\Omega_f^*$ is continuous and one to one and $Ran\,\Omega_f^*$ is dense in $\mathcal H$. Define the operator 
$M_F:\mathcal H\mapsto\mathcal H$ by
\begin{equation*}
 M_F:=\Omega_f^*\Omega_f\,.
\end{equation*}
By (II) above and the fact that $\Omega_f$, $\Omega_f^*$ are quasi-affine we get that $M_F$ is a quasi-affine mapping from
$\mathcal H$ into $\mathcal H$. Therefore $M_F$ is continuous and injective and $Ran\,M_F$ is dense in $\mathcal H$.
Obviously $M_F$ is symmetric and, since $\Omega_f$ and $\Omega_f^*$ are bounded, then
$Dom\, M_F=\mathcal H$ and we conclude that $M_F$ is self-adjoint. Since $\Omega_f$ and $\Omega_f^*$ are both contractive
then $M_F$ is contractive. In fact, it is shown in \cite{S3} that $\Vert M_F\Vert=1$. 
\smallskip
\par{\bf Remark:} It is to be noted that the operator $M_F$ already appears in reference \cite{S3} in a slightly different context. 
Indeed, $M_F$ is identical to the inverse $T_F^{-1}$ of the operator $T_F$ called the \emph{time observable} in that paper. 
\smallskip
\par Taking the adjoint of Eq. (\ref{basic_intertwining_rel_eqn}) we obtain
\begin{equation}
\label{adj_basic_intertwining_rel_eqn}
 U(-t)\Omega^*_f g=\Omega^*_f(T_u(t))^*g,\qquad t\geq 0,\ \ g\in\mathcal H^2_+(\mathbb R),
 \qquad t\geq 0,\ \ g\in\mathcal H^2_+(\mathbb R)\,,
\end{equation}
we obtain from Eqns. (\ref{basic_intertwining_rel_eqn}) and  (\ref{adj_basic_intertwining_rel_eqn}) an expression for the
Heisenberg evolution of $M_F$
\begin{equation*}
 U(-t)M_FU(t)=U(-t)\Omega_f^*\Omega_f U(t)=\Omega_f^* (T_u(t))^*T_u(t)\Omega_f\,.
\end{equation*}
For any $\psi\in\mathcal H$ we then get
\begin{equation*}
 (\psi,U(-t)M_FU(t)\psi)=(\psi,\Omega_f^*(T_u(t))^*T_u(t)\Omega_f\psi)
 =\Vert T_u(t)\Omega_f\psi\Vert^2,\qquad t\geq 0,\ \ \psi\in\mathcal H\,.
\end{equation*}
The fact that $M_F$ is a Lyapounov variable, i.e., the validity of Eqns. (\ref{M_F_lyapounov_var_eqn}) and (\ref{M_F_to_zero_eqn})
then follows immediately from Eqns. (\ref{T_semigroup_eqn}) and (\ref{T_to_zero_eqn}).
\par We are left with the task of showing that $M_F$ can be expressed in the form given by Eq. (\ref{M_F_expression_eqn}).
For this we need a more explicit expression for the map $\Omega_f$. It follows from assumptions (a)-(c) in Theorem
\ref{exist_lyapounov_var_thm} that there exists a unitary mapping $U:\mathcal H\mapsto L^2(\mathbb R^+)$ of $\mathcal H$ into 
its spectral representation on the spectrum of $H$ (energy representation for $H$). The energy representation
is obtained by finding a complete set of improper eigenvectors $\{\phi_E\}_{E\in\mathbb R^+}$ of $H$, corresponding to the
(by assumption absolutely continuous) spectrum of $H$. Using the Dirac notation 
$\{\phi_E\}_{E\in\mathbb R^+}\equiv\{ \vert E\rangle\}_{E\in\mathbb R^+}$ we have
\begin{equation}
\label{H_to_energy_rep_eqn}
 (U\psi)(E)=\langle E\vert\psi\rangle=\psi(E),\qquad E\in\mathbb R^+,\ \ \psi\in\mathcal H\,.
\end{equation}
the inverse of $U$ is given by
\begin{equation}
\label{energy_rep_to_H_eqn}
 U^*f=\int_0^\infty dE\, \vert E\rangle\, \psi(E),\qquad \psi\in L^2(\mathbb R^+)\,.
\end{equation}
Let $P_{\mathbb R^+}:L^2(\mathbb R)\mapsto L^2(\mathbb R)$ be the orthogonal projection in $L^2(\mathbb R)$ on the subspace 
of functions supported on $\mathbb R^+$ and define the inclusion map $I: L^2(\mathbb R^+)\mapsto L^2(\mathbb R)$ by
\begin{equation*}
 (If)(\sigma)=\begin{cases}
                  f(\sigma), & \sigma\geq 0\\
                             0, & \sigma< 0\\
                  \end{cases}
                  ,\qquad \sigma\in\mathbb R\,.
\end{equation*}
Then the inverse $I^{-1}: P_{\mathbb R^+}L^2(\mathbb R)\mapsto L^2(\mathbb R^+)$ is well defined on 
$P_{\mathbb R^+}L^2(\mathbb R)$. Let $\theta:\mathcal H^2_+(\mathbb R)\mapsto L^2(\mathbb R^+)$ be given by
\begin{equation*}
 \theta f=I^{-1} P_{\mathbb R^+} f,\qquad f\in\mathcal H^2_+(\mathbb R)\,.
\end{equation*}
By a theorem of Van-Winter  \cite{V} $\theta$ is a contractive quasi-affine mapping of $\mathcal H^2_+(\mathbb R)$ into 
$L^2(\mathbb R^+)$. The adjoint map $\theta^*: L^2(\mathbb R^+)\mapsto\mathcal H^2_+(\mathbb R)$ is then also a contractive
quasi-affine map. An explicit expression for $\theta^*$ is given in \cite{S1,S2}
\begin{equation*}
 \theta^* f=P_+ I f,\qquad f\in L^2(\mathbb R^+)\,.
\end{equation*}
It is shown in \cite{S1,S2} that the maps $\Omega_f$ and $\Omega_f^*$ are given by
\begin{equation*}
 \Omega_f=\theta^*U,\qquad \Omega_f^*=U^*\theta\,.
\end{equation*}
From the definition of the embedding map $I$ and the expression for $P_+$ in Eq. (\ref{hardy_projections_eqn}) we get
\begin{equation}
\label{theta_star_explicit_eqn}
 (\theta^* f )(\sigma)=\frac{-1}{2\pi i}\int_0^\infty dE\, \frac{1}{\sigma-E+i0^+} f(E),\qquad \sigma\in\mathbb R,\ f\in L^2(\mathbb R^+)\,.
\end{equation}
Combining Eqns. (\ref{H_to_energy_rep_eqn}), (\ref{energy_rep_to_H_eqn}) and Eq. (\ref{theta_star_explicit_eqn}) we finally obtain
\begin{equation*}
 M_F=\Omega_f^*\Omega_f=\frac{-1}{2\pi i}\int_0^\infty dE'\int_0^\infty dE\, \vert E' \rangle\frac{1}{E'-E+i0^+}\langle E\vert\,.
\end{equation*}
\par\hfill$\blacksquare$
\par\smallskip\
\par We now proceed to the proof of the second main result of this paper:
\bigskip
\par{\bf Proof of Theorem \ref{lambda_transform_thm}}
\smallskip
\par Since $M_F$ is a bounded positive operator its positive square root $\Lambda_F$ is well defined and unique \cite{R}
and we set $\Lambda_F:=M_F^{1/2}=(\Omega_f^*\Omega_f)^{1/2}$. Moreover, since 
\begin{equation*}
 M_F\mathcal H=\Lambda_F\Lambda_F \mathcal H\subseteq\Lambda_F\mathcal H\,,
\end{equation*}
and since $Ran\,M_F$ is dense in $\mathcal H$ we conclude that $Ran\,\Lambda_F$ is dense in $\mathcal H$. Furthermore,
since $M_F=\Lambda_F^2$ is one to one then $\Lambda_F$ must also be one to one. We can summarize the findings above by 
stating that the fact that $M_F$ is positive, one to one and quasi-affine implies the same properties for $\Lambda_F$.
Since $M_F$ is contractive and since for every $\psi\in\mathcal H$ we have $(\psi,M_F\,\psi)=\Vert \Lambda_F\psi\Vert^2$ we 
conclude that $\Lambda_F$ is also contractive. 
\par Define a mapping $\tilde R: \mathscr D(\tilde R)\mapsto\mathcal H^2_+(\mathbb R)$ with   
$\mathscr D(\tilde R)\subseteq\mathcal H$ and
\begin{equation}
\label{R_tilde_def_eqn}
 \tilde R:=\Omega_f (\Omega_f^*\Omega_f)^{-1/2}=\Omega_f \Lambda_F^{-1}\,.
\end{equation}
Obviously $\mathscr D(\tilde R)\supseteq Ran\,\Lambda_F\supset Ran\,M_F$ so that $\tilde R$ is defined on a dense set in 
$\mathcal H$. For any $f\in\mathscr D(\tilde R)$ we have
\begin{multline}
\label{R_tilde_isometric_eqn}
 \Vert\tilde R f\Vert^2=(\tilde R f,\tilde R f)=(\Omega_f(\Omega_f^*\Omega_f)^{-1/2}f,\Omega_f(\Omega_f^*\Omega_f)^{-1/2}f)=\\
 =((\Omega_f^*\Omega_f)^{-1/2}f,\Omega_f^*\Omega_f(\Omega_f^*\Omega_f)^{-1/2}f)
  =((\Omega_f^*\Omega_f)^{-1/2}f,(\Omega_f^*\Omega_f)^{1/2}f)=\Vert f\Vert^2\,,
\end{multline}
hence $\tilde R$ is isometric on a dense set in $\mathcal H$ and can be extended to an isometric map
$R:\mathcal H\mapsto\mathcal H^2_+(\mathbb R)$ such that 
\begin{equation*}
 R^*R=I_{\mathcal H}\,.
\end{equation*}
From Eq. (\ref{R_tilde_isometric_eqn}) we see that on the dense set $Ran\,\tilde R\subset\mathcal H^2_+(\mathbb R)$ we have 
\begin{equation*}
 \tilde R^* f=(\Omega_f^*\Omega_f)^{-1/2}\Omega_f^* f=\Lambda_F^{-1} \Omega_f^* f,\qquad f\in Ran\,\tilde R
\end{equation*}
and the adjoint $R^*$ of $R$ is an extension of $\tilde R^*$ to $\mathcal H^2_+(\mathbb R)$. Note that the definition of $\tilde R$
implies that $Ran\,\tilde R\subseteq Ran\,\Omega_f$. Hence, for any $g\in Ran\,\Omega_f$ we have
\begin{equation}
\label{R_tilde_star_eqn}
 \tilde R^* g=(\Omega_f^*\Omega_f)^{-1/2}\Omega_f^*g
 =(\Omega_f^*\Omega_f)^{-1/2}\Omega_f^*\Omega_f\Omega_f^{-1}g
 =(\Omega_f^*\Omega_f)^{1/2}\Omega_f^{-1}=\Lambda_f\Omega_f^{-1}g\,.
\end{equation}
Thus on the dense set $Ran\,\Omega_f\subset \mathcal H^2_+(\mathbb R)$ we have
\begin{equation*}
 R\,R^*g=[\Omega_f(\Omega_f^*\Omega_f)^{-1/2}] [(\Omega_f^*\Omega_f)^{1/2}\Omega_f^{-1}] g=g
\end{equation*}
and by continuity we obtain 
\begin{equation*}
 R\,R^*=I_{\mathcal H^2_+(\mathbb R)}
\end{equation*}
and hence $R:\mathcal H\mapsto\mathcal H^2_+(\mathbb R)$ is, in fact, a unitary map. 
\par Now define
\begin{equation*}
 Z(t):=\Lambda_F U(t)\Lambda_F^{-1},\qquad t\geq 0\,.
\end{equation*}
Obviously, $Z(t)$ is well defined on $Ran\,\Lambda_F$ for any $t\geq 0$. Moreover, using the definition of $\tilde R$ from 
Eq. (\ref{R_tilde_def_eqn}) and Eqns. (\ref{R_tilde_star_eqn}), (\ref{basic_intertwining_rel_eqn}) we get
\begin{multline*}
 RZ(t)R^* g=\Omega_f\Lambda_F^{-1} Z(t)\Lambda_F\Omega_f^{-1} g
 =[\Omega_f\Lambda_F^{-1}]\,[\Lambda_F U(t)\Lambda_F^{-1}]\,[\Lambda_F\Omega_f^{-1}]g=\\
 =\Omega_f U(t)\Omega_f^{-1}g=T_u(t)g,\quad t\geq 0,\ \ g\in Ran\,\Omega_f\subset\mathcal H^2_+(\mathbb R)\,.
\end{multline*}
Then on the dense subset $R^*\Lambda_F\mathcal H\subset\mathcal H$ we have $Z(t)=R^*T_u(t)R$ and since $R$ and
$T_u(t)$ are bounded we are able by continuity to extend the domain of definition of $Z(t)$ to all of $\mathcal H$ and obtain
\begin{equation}
\label{Z_unitary_transform_eqn}
 RZ(t)R^*=T_u(t),\quad Z(t)=R^*T_u(t)R,\ \ t\geq 0\,.
\end{equation}
From the unitarity of $R$, the fact that $\{T_u(t)\}_{t\in\mathbb R^+}$ is a continuous, strongly contractive, one parameter
semigroup and Eqns. (\ref{T_semigroup_eqn}), (\ref{T_to_zero_eqn}) we conclude that $\{Z(t)\}_{t\in\mathbb R^+}$ is a
continuous, strongly contractive, one parameter semigroup and Eqns. (\ref{Z_semigroup_prop_eqn}) and 
(\ref{Z_intertwining_eqn}) hold. \hfill$\blacksquare$
\par\bigskip\ 
\par{\bf Proof of Theorem \ref{T_observable_thm}:}
\smallskip
\par The main results of Theorem \ref{T_observable_thm} are consequences of the following lemma:
\begin{lemma}
\label{hardy_semigroup_projections_eqn}
For every $t\geq 0$ the operator $T_u(t):\mathcal H^2_+(\mathbb R)\mapsto\mathcal H^2_+(\mathbb R)$  is isometric and
we have 
\begin{equation}
\label{T_star_range_eqn}
 Ran\,(T_u(t))^*=(Ker\,T_u(t))^\perp
\end{equation}
and
\begin{equation*}
 T_u(t)(T_u(t))^*=I_{\mathcal H^2_+(\mathbb R)},\quad t\geq 0\,.
\end{equation*}
Furthermore, if $\hat P_{t]}:\mathcal H^2_+(\mathbb R)\mapsto\mathcal H^2_+(\mathbb R)$ is the orthogonal projection on 
$Ker\,T_u(t)$ and $\hat P_{[t}$ is the orthogonal projection on $(Ker\,T_u(t) )^\perp$ then 
\begin{equation*}
 \hat P_{t]}=[T_u(t),(T_u(t))^*],\quad t\geq 0
\end{equation*}
and
\begin{equation*}
 \hat P_{[t}=(T_u(t))^*T_u(t),\quad t\geq 0\,.
\end{equation*} 
Moreover, we have
\begin{equation}
\label{P_t_right_product_eqn}
 \hat P_{t_1]}\hat P_{t_2]}=\hat P_{t_1]},\ \ t_2\geq t_1\geq 0,\qquad Ran\,\hat P_{t_1]}\subset Ran\,\hat P_{t_2]},\quad t_2>t_1
\end{equation}
and 
\begin{equation*}
 \hat P_{0]}=0,\qquad \lim_{t\to\infty} \hat P_{t]}=I_{\mathcal H^2_+(\mathbb R)}\,.
\end{equation*}
\hfill$\square$
\end{lemma}
\par{\bf Proof of Lemma \ref{hardy_semigroup_projections_eqn}:}
\smallskip
\par Recall that $T_u(t)f=P_+u(t)f$ for $t\geq 0$ and $f\in\mathcal H^2_+(\mathbb R)$. Since $\mathcal H^2_+(\mathbb R)$ is 
stable under $u^*(t)=u(-t)$ for $t\geq 0$, i.e., $u(-t)\mathcal H^2_+(\mathbb R)\subset\mathcal H^2_+(\mathbb R)$ (as one can 
see, for example, by using the Paley-Wiener theorem \cite{PW}), we find that for any $f,g\in\mathcal H^2_+(\mathbb R)$ we 
have
\begin{equation*}
 (g,T_u(t)f)=(g,P_+u(t)f)=(u(-t)g,f)=(P_+u(-t)g,f)=(T_{u^*}(t)g,f)=((T_u(t))^*g,f)\,.
\end{equation*}
Therefore
\begin{equation}
\label{T_star_expression_eqn}
 (T_u(t))^*g=u(-t)g,\quad t\geq 0,\ \ g\in\mathcal H^2_+(\mathbb R)\,.
\end{equation}
Since $u(-t)$ is unitary on $L^2(\mathbb R)$ Eq. (\ref{T_star_expression_eqn}) implies that $(T_u(t))^*$ is isometric
on $\mathcal H^2_+(\mathbb R)$. The same equation implies also that
\begin{equation}
\label{T_T_star_unit_eqn}
 (T_u(t)(T_u(t))^*f=P_+u(-t)u(t)f=P_+f=f,\quad t\geq 0,\ \ f\in\mathcal H^2_+(\mathbb R)\,.
\end{equation}
Consider now the operator $A(t):=(T_u(t))^*T_u(t)$ for $t\geq 0$. Since $Dom\,T_u(t)=\mathcal H^2_+(\mathbb R)$ we have
that $A(t)$ is self-adjoint. In addition Eq. (\ref{T_T_star_unit_eqn} implies that 
\begin{equation*}
 (A(t))^2=[(T_u(t))^*T_u(t)][(T_u(t))^*T_u(t)]=(T_u(t))^*T_u(t)=A(t),\quad t\geq 0\,,
\end{equation*} 
so that $A(t)$ is an orthogonal projection in $\mathcal H^2_+(\mathbb R)$. Of course, for any $u\in Ker\,T_u(t)$ we have
$A(t)u=0$, hence $Ran\,A(t)\subseteq Ker\,T_u(t)$. Assume that there is some 
$v\in(Ran\,A(t))^\perp\cap(Ker\,T_u(t))^\perp$ with $v\not= 0$. Then we must have $A(t)v=0$, but since $T_u(t)v\not=0$
and since $(T_u(t))^*$ is an isometry we obtain a contradiction. Therefore $Ran\,A(t)=(Ker\,T_u(t))^\perp$ and
$\hat P_{[t}=A(t)=(T_u(t))^*T_u(t)$. Taking into account Eq. (\ref{T_T_star_unit_eqn}) we obtain also
$\hat P_{t]}=I-\hat P_{[t}=[T_u(t),(T_u(t))^*]$.
\par To prove Eq. (\ref{T_star_range_eqn}) we note that since $(T_u(t))^*$ is isometric its range is a close subspace of 
$\mathcal H^2_+(\mathbb R)$ and, moreover, $(Ran\,(T_u(t))^*)^\perp\supseteq Ker\,T_u(t)$. This is a result of the fact that if
$u\in Ker\,T_u(t)$ then
$(u,(T_u(t))^*v)=(T_u(t)u,v)=0$, $\forall v\in\mathcal H^2_+(\mathbb R)$. On the other hand, if $u$ is orthogonal to 
$Ran\,(T_u(t))^*$ i.e., $u$ is such that $(u,(T_u(t))^*v)=0$, $\forall v\in\mathcal H^2_+(\mathbb R)$ then 
$(T_u(t)u,v)=0$, $\forall v\in\mathcal H^2_+(\mathbb R)$ so that $u\in Ker\,T_u(t)$ and we get that 
$(Ran\,(T_u(t))^*)^\perp\subseteq Ker\,T_u(t)$.
\par In order to verify the validity the first equality in Eq. (\ref{P_t_right_product_eqn}) we use the semigroup property of 
$\{T_u(t)\}_{t\in\mathbb R}$ and Eq. (\ref{T_T_star_unit_eqn}). For $t_1\leq t_2$ we get
\begin{multline*}
 \hat P_{t_1]}\hat P_{t_2]}=(I-\hat P_{[t_1})(I-\hat P_{[t_2})
 = I-\hat P_{[t_1}-\hat P_{[t_2}+(T_u(t_1))^*T_u(t_1)(T_u(t_2))^*T_u(t_2)=\\
 =I-\hat P_{[t_1}-\hat P_{[t_2}+(T_u(t_1))^*(T_u(t_2-t_1))^*T_u(t_2)
 =I-\hat P_{[t_1}-\hat P_{[t_2}+(T_u(t_2))^*T_u(t_2)=\\
  =I-\hat P_{[t_1}-\hat P_{[t_2}+\hat P_{[t_2}=\hat P_{t_1]}\,.
\end{multline*}
We need to show also that $Ker\,T_u(t_1)\subset Ker\,T_u(t_2)$ for $t_2>t_1$. Note that since for $t_2>t_1$ we have
$T_u(t_2)=T_u(t_2-t_1)T_u(t_1)$ and since $(T_u(t))^*$ is isometric on $\mathcal H^2_+(\mathbb R)$ then it is enough
to show that $Ker\,T_u(t)\not=\{0\}$ for every $t>0$. If this condition is true and if $f\in Ker\,T_u(t_2-t_1)$
we just set $g=(T_u(t_1))^*f$ and we get that 
\begin{equation*}
 T_u(t_1)g=T_u(t_1)(T_u(t_1))^*f=f
\end{equation*}
and
\begin{equation*}
 T_u(t_2)g=T_u(t_2)(T_u(t_1))^*f=T_u(t_2-t_1)f=0\,.
\end{equation*}
In order to show that $Ker\,T_u(t)\not=\{0\}$ for every $t>0$ we exhibit a state belonging to this kernel. Indeed one may easily
check that for a complex constant $\mu$ such that $Im\,\mu<0$ and for $t_0>0$ the function
\begin{equation*}
 f(\sigma)=\frac{1}{\sigma-\mu}\left[1-e^{i\sigma t_0}e^{-i\mu t_0}\right],\quad \sigma\in\mathbb R\
\end{equation*}
is such that $f\in Ker\,T_u(t)\subset\mathcal H^2_+(\mathbb R)$ for every $t\geq t_0>0$.
\par Finally, it is immediate that $\hat P_{0]}=0$ and, moreover, since for every $f\in\mathcal H^2_+(\mathbb R)$ we have
$\Vert \hat P_{[t} f\Vert^2=(f,\hat P_{[t} f)=(f,(T_u(t))^*T_u(t)f)=\Vert T_u(t) f\Vert^2$ then $s-\lim_{t\to\infty}\hat P_{[t}=0$ and
hence $s-\lim_ {t\to\infty}\hat P_{t]}=I_{\mathcal H^2_+(\mathbb R)}$.\hfill$\blacksquare$
\par\bigskip\
\par For $t\geq 0$ define $P_{t]}:=R^*\hat P_{t]}R$ and $P_{[t}:=R^*\hat P_{[t}R=I_{\mathcal H}-P_{t]}$. Combining Theorem 
\ref{hardy_semigroup_projections_eqn} and Eq. (\ref{Z_unitary_transform_eqn}) and taking into account
the unitarity of the mapping $R$ we conclude that there exists families $\{P_{t]}\}_{t\in\mathbb R^+}$,
$\{P_{[t}\}_{t\in\mathbb R^+}$,  of orthogonal projections in $\mathcal H$ such that $P_{t]}+P{[t}=I_{\mathcal H}$ and 
\begin{equation*}
 Ran\,P_{t]}=Ker\,Z(t),\quad Ran\,P_{[t}=(Ker\,Z(t))^\perp,\quad t\geq 0\,,
\end{equation*}
\begin{equation*}
 P_{t]}=[Z(t),Z^*(t)],\quad t\geq 0\,,
\end{equation*}
\begin{equation*}
 P_{[t}=Z^*(t)Z(t),\quad t\geq 0\,,
\end{equation*} 
\begin{equation}
\label{P_Z_right_product_eqn}
 P_{t_1]} P_{t_2]}=P_{t_1]},\ \ t_2\geq t_1\geq 0,\qquad Ran\, P_{t_1]}\subset Ran\, P_{t_2]},\quad t_2>t_1
\end{equation}
and 
\begin{equation}
\label{P_Z_limits_eqn}
 P_{0]}=0,\qquad \lim_{t\to\infty} P_{t]}=I_{\mathcal H}\,.
\end{equation}
In addition we have
\begin{equation*}
 Ran\,(Z^*(t))=(Ker\,Z(t))^\perp
\end{equation*}
and
\begin{equation*}
 Z(t)Z^*(t)=I_{\mathcal H},\quad t\geq 0\,.
\end{equation*}
Eqns. (\ref{P_Z_right_product_eqn}), (\ref{P_Z_limits_eqn}) imply that it is possible to construct from the family
$\{P_{t]}\}_{t\in\mathbb R^+}$ of orthogonal projections a spectral family of a corresponding self-adjoint operator. First define for
intervals
\begin{equation*}
\mu_T(A)=\begin{cases}
            P_{b]}-P_{a]},                 & A=(a,b]\,.\\
            P_{b]}-P_{(a-0^+)]},         & A=[a,b]\,,\\
            P_{(b-0^+)]}-P_{a]},         & A=(a,b)\,,\\
            P_{(b-0^+)]}-P_{(a-0^+)]}, & A=[a,b)\,,\\
             \end{cases}
\end{equation*}
where $b>a> 0$ (and with $P_{(a-0^+)]}$ replaced by $P_{0]}$ for $a=0$), and then extend $\mu_T$ to the Borel $\sigma$-algebra 
of $\mathbb R^+$. Following the definition of the
spectral measure $\mu_T: \mathscr B(\mathcal H)\mapsto \mathcal P(\mathcal H)$ we subsequently are able to define a
self-adjoint operator $T:\mathscr D(T)\mapsto \mathcal H$ via
\begin{equation*}
 T:=\int_0^\infty t\,d\mu_T(t)\,.
\end{equation*}
By construction it is immediate that $T$ has the properties listed in Theorem \ref{T_observable_thm}. For example, we have
\begin{equation*}
 \mu([0,t])\mathcal H=(P_{t]}-P_{0]})\mathcal H=P_{t]}\mathcal H=[Z^*(t),Z(t)]\mathcal H
\end{equation*}
and
\begin{equation*}
 \mu([t,\infty)\mathcal H=\lim_{t'\to\infty}(P_{t']}-P_{t]})\mathcal H=(I_{\mathcal H}-P_{t]})\mathcal H
 =P_{[t}\mathcal H=Z^*(t)Z(t)\mathcal H\,.
\end{equation*}
\hfill$\blacksquare$
\par\smallskip\ 
\par This concludes the proofs of the three main results of this paper.
%
%
%
\section{Summary}
\label{summary}
\par The Misra, Prigogine and Courbage theory of classical and quantum microscopic irreversibility is based on the
notion of Lyapounov variables. It is known from the Poincare'-Misra theorem that in the classical theory Lyapounov variables
corresponding to non-equilibrium entropy cannot be associated with phase-space functions. In fact, it was shown by Misra that in
Koopman's Hilbert space formulation of classical mechanics an operator corresponding to a Lyapounov variable
cannot commute with all of the operators of multiplication by phase space functions. In quantum theory it was shown by 
Misra, Prigogine and Courbage that under assumptions (i)-(v) in Section \ref{introduction} there does not exist a Lyapounov variable
as an operator in the Hilbert space $\mathcal H$ corresponding to the given quantum mechanical problem. The solution to this 
problem found by Misra, Prigogine and Courbage is to turn to the Liouvillian representation of quantum mechanics and define
the Lyapounov variable as a super operator on the space of density matrices. Then, under the assumption that the Hamiltonian $H$
of the problem has absolutely continuous spectrum $\sigma(H)=\sigma_{ac}(H)=\mathbb R^+$ it is possible to carry out the 
program, define a Lyapounov variable as a super operator and find a non-unitary $\Lambda$-transformation to an irreversible
representaion of the quantum dynamics.
\par In the present paper it is shown that if one relaxes conditions (i)-(v) in Section \ref{introduction} then, under the
same assumptions on the spectrum of the Hamiltonian made by Misra, Prigogine and Courbage, it is possible to construct a
Lyapounov variable for the original Schr\"odinger evolution $U(t)=\exp(-i Ht)$, $t\geq 0$ as an operator in the Hilbert space 
$\mathcal H$ of the given quantum mechanical problem without resorting to work in Liouville space and defining a Lyapounov
variable as a super operator acting on density matrices. The method of proof of the existence of a Lyapounov variable is 
constructive and an explicit expression for such an operator is given in the form of Eq. (\ref{M_F_expression_eqn}). Moreover,
it is shown that a $\Lambda$-transformation to an irreversible representation of the dynamics can be defined also in this case.
Finally, it is demonstrated that the irreversible representation of the dynamics is the natural representation of the flow of time
in the system in the sense that there exists a positive, semibounded operator $T$ in $\mathcal H$ such that if $\mu_T$ is the 
spectral projection valued measure of $T$ then for each $t\geq 0$ the spectral projections $P_{t]}=\mu_T([0,t))$ and 
$P_{[t}=(I_{\mathcal H}-P_{t]})=\mu_T([t,\infty))$ split the Hilbert space $\mathcal H$ into the direct sum of a past subspace
$\mathcal H_{t]}$ and a future subspace $\mathcal H_{[t}$
\begin{equation*}
 \mathcal H=\mathcal H_{t]}\oplus\mathcal H_{[t},\quad \mathcal H_{t]}=P_{t]}\mathcal H,
 \quad \mathcal H_{[t}=P_{[t}\mathcal H,\ \ t\geq 0
\end{equation*}
such that, as its name suggests, the past subspace $\mathcal H_{t]}$ at time $t\geq 0$ does not enter into the calculation of
any matrix element of any observable for all times $t' > t\geq 0$, i.e., at time $t$ it already belongs to the past. Put differently, in
the irreversible representation the operator $T$ provides us with a super selection rule separating past and future as there is no 
observable for the system that can connect the past subspace to the future subspace and all matrix elements and expectation 
values for $t'>t>0$ are, in fact, calculated in the future subspace $\mathcal H_{[t}$.
\par\bigskip\ 
\par\centerline{\Large{\bf Acknowledgements}}
\smallskip
\par Research supported by ISF under Grant No. 1282/05 and by the Center for Advanced Studies in Mathematics at 
Ben-Gurion University.
\bigskip

\end{document}